\documentclass[preprint,12pt]{elsarticle}
\usepackage{geometry}                
\usepackage{graphicx}
\usepackage{amssymb}
\usepackage{epstopdf}

\usepackage{amssymb}

\journal{Physics Letters B}

\begin{document}

\begin{frontmatter}



\title{Low-energy multipole response in nuclei at finite temperature}


\author[1,2]{Y. F. Niu}
\author[1]{N. Paar}
\author[1]{D. Vretenar}
\author[2,3]{J. Meng}

\address[1]{Physics Department, Faculty of Science, University of Zagreb, Croatia}
\address[2]{State Key Laboratory for Nuclear Physics and Technology, School of Physics, Peking University, Beijing 100871, China}
\address[3]{School of Physics and Nuclear Energy, Beihang University, Beijing 100083, China}

\begin{abstract}
The multipole response of nuclei at temperatures 
$T= 0 -2$ MeV is studied using a self-consistent finite-temperature 
RPA (random phase approximation) based on relativistic 
energy density functionals. Illustrative calculations are performed 
for the isoscalar monopole and isovector dipole modes and, 
in particular, the evolution of low-energy excitations with temperature is 
analyzed, including the modification of pygmy structures. Both for the 
monopole and dipole modes, in the temperature range $T= 1 -2$ MeV 
additional transition strength appears at low energies because of thermal 
unblocking of single-particle orbitals close to the Fermi level. A concentration 
of dipole strength around 10 MeV excitation energy is predicted in  $^{60,62}$Ni, 
where no low-energy excitations occur at zero temperature. 
The principal effect of finite temperature on low-energy strength that is 
already present at zero temperature, 
e.g. in $^{68}$Ni and $^{132}$Sn, is the spreading of this structure to even 
lower energy and the appearance of states that correspond to
thermally unblocked transitions.
\end{abstract}

\begin{keyword}
hot nuclei \sep pygmy resonances \sep giant resonances \sep random phase approximation
\PACS 21.10.Gv \sep 21.30.Fe \sep 21.60.Jz \sep 24.30.Cz


\end{keyword}

\end{frontmatter}






The multipole response of nuclei at energies close to the neutron emission threshold, 
and the possible occurrence of exotic low-energy modes of excitation, present a rapidly 
growing field of research \cite{PVKC.07}. For instance, in nuclei with a pronounced 
neutron excess, the phenomenon of  pygmy dipole resonance (PDR), a low-energy mode 
that  corresponds to oscillations of a weakly-bound neutron skin against the 
isospin saturated proton-neutron core, has been analyzed in a number of 
experimental and theoretical studies. In particular, recent $(\gamma,\gamma')$ and 
$(\alpha,\alpha'\gamma)$ experiments \cite{Sav.06,Sch.08,Oze.08} have disclosed 
interesting features of the concentration of low-energy electric-dipole excitations. 
Exotic nuclear dynamics has also been explored in experiments with radioactive ion beams,
extending the studies of low-lying dipole response toward nuclei with more pronounced 
neutron excess~\cite{Adr.05,Wie.09}.

Features of PDR in medium-heavy 
and heavy nuclei place constraints on the nuclear matter symmetry 
energy, and can also provide information on the size
of the neutron skin~\cite{Pie.06,Kli.07}. A concentration of 
electric-dipole strength at low energy 
is interesting not only as a unique structure phenomenon, but it 
could also play an important role in astrophysical processes. For 
instance,  it could affect neutron capture rates in r-process 
nucleosynthesis \cite{Gor.98,Lit.09}. Reactions relevant for the 
r-process take place in stellar environment at finite temperature. 
Calculations of neutron capture rates on neutron-rich nuclei
have typically included temperature only as a parameter of the Lorentzian 
function used to fold  the E1 strength distribution~\cite{Gor.98,Gor.02}.
It would, of course, be useful to perform these calculations using a 
microscopic framework that  provides a consistent description of
nuclear ground-state properties and excitations at finite temperature. 
In particular, temperature induced modifications of low-energy dipole
strength close to the neutron separation energy, e.g 
the thermal unblocking of additional transitions, should be 
described in a consistent theoretical framework. 

In this work we introduce a fully self-consistent finite-temperature 
RPA (random phase approximation) framework based on relativistic 
energy density functionals, and explore the evolution of multipole 
response as a function of temperature. The goal is to analyze how 
low-energy excitations evolve with temperature, the modification 
of the structure of PDR, and the possible occurrence of excitation modes 
that are not present in nuclei at zero temperature.

The multipole response of hot nuclei has been studied using 
a variety of theoretical approaches, including e.g. the RPA 
based on schematic interactions~\cite{Civ.84,Vau.84,Bes.84},
linear response theory~\cite{Fab.83,Rin.84}, Hartree-Fock plus RPA 
with Skyrme effective interactions~\cite{Sag.84},  RPA models 
extended with the inclusion of collision terms~\cite{Lac.98}. 
The decay of hot nuclei and thermal damping of giant resonances 
have also been explored in numerous studies, e.g. in 
Refs.~\cite{Gal.85,Bor.86,Dan.96,OBA.96,Dan.97,Lar.99,Sto.04},
as well as sum rules at finite temperature~\cite{Mey.83,Bar.85}. 
Finite temperature Hartree-Fock Bogoliubov (HFB) theory \cite{Goo.81} 
has been used as the basis for the corresponding  quasiparticle 
RPA~\cite{Som.83,Kha.04}. Experimental studies of giant resonances 
in hot nuclei are carried out by measuring high energy $\gamma$-rays 
from the decay of resonant states built on excited states of target 
nuclei~\cite{Bra.89,Tve.96,Pie.96,Bau.98}.

The self-consistent finite temperature 
relativistic random phase approximation (FTRRPA) is
formulated in the single-nucleon basis of the relativistic mean-field (RMF) 
model at finite temperature (FTRMF). This model has been introduced in 
Ref.~\cite{Gam.00}, based on the nonlinear Lagrangian NL3 \cite{LKR.97}.
In the theoretical framework that we use in the present analysis 
relativistic effective interactions 
are implemented self-consistently, i.e., both the FTRMF equations and the 
matrix equations of FTRRPA are based on the same relativistic
energy density functional with medium-dependent meson-nucleon 
couplings~\cite{NVFR.02}. Of course, the 
description of open-shell nuclei necessitates a consistent treatment
of pairing correlations like, for instance, in the finite temperature 
HFB+QRPA framework~\cite{Som.83}.
However, in nuclei the phase transition from a superfluid to a normal
state occurs at temperatures $T \approx 0.5-1$ MeV~\cite{Goo.81, Kha.04}, 
whereas for temperatures above $T \approx 4$ MeV
contributions from states in the continuum become
large, and additional subtraction schemes have to be implemented to remove
the contributions of the external nucleon gas~\cite{Bon.84}. Therefore,
we consider the temperatures in the range $T=1-2$ MeV, for which 
the FTRMF plus FTRRPA should provide an accurate description of the 
evolution of multipole response.

The FTRRPA represents the small amplitude limit of the time-dependent
relativistic mean-field theory at finite temperature.
Starting from the response of the time-dependent density matrix $\hat{\rho}(t)$ to
an external field $\hat{f}(t)$, the equation of motion for the density operator
reads 
\begin{equation}
\label{EOM}
 i\partial_t \hat{\rho} = [ \hat{h}[ \hat{\rho}]+ \hat{f}(t), \hat{ \rho} ] \; .
\end{equation}
In the small amplitude limit the density matrix is expanded to linear order
\begin{equation}
 \hat{\rho} (t)= \hat{\rho}^{0}+ \delta \hat{\rho}(t),
 \end{equation}
where $\hat{\rho}^{0}$ denotes the stationary ground-state density.
\begin{equation}
\rho^{0}_{kl}= \delta_{kl}n_k = \left\{ \begin{array}{cl} [1+exp(\frac{\epsilon_k - \mu}{kT})]^{-1} & 
\mbox{for states in the Fermi sea (index $k, l$) }\\
0 & \mbox{for unoccupied states in the Dirac sea (index $\alpha$)} \end{array}
\right.
 \end{equation}
and includes the thermal occupation factors of single-particle states $n_k$.
The resulting FTRRPA equations read
\begin{equation}
   \left( \begin{array}{cc} A & B \\ -B^* & -A^* \end{array} \right)
   \left( \begin{array}{c} X \\ Y \end{array} \right)
   = \hbar\omega \left( \begin{array}{c} X \\ Y \end{array} \right) \;,
\end{equation}
where 
\begin{equation}
   A = \left( \begin{array}{cc} (\epsilon_m - \epsilon_i)
   \delta_{ii'} \delta_{mm'} &  \\
   & (\epsilon_\alpha - \epsilon_i) \delta_{\alpha \alpha'}
   \delta_{ii'} \end{array} \right)
   + \left( \begin{array}{cc} (n_{i'} - n_{m'})V_{mi'im'} & n_{i'} V_{mi'i\alpha'} \\
   (n_{i'} - n_{m'})V_{\alpha i' i m'}  &n_{i'} V_{\alpha i' i \alpha'} \end{array}
   \right) \;, 
  \end{equation}
  and
\begin{equation}
   B =\left( \begin{array}{cc} (n_{i'} - n_{m'})V_{mm'ii'} & n_{i'}V_{m\alpha'ii'} \\
    (n_{i'} - n_{m'})V_{\alpha m' i i'}  & n_{i'}  V_{\alpha \alpha' i i' } \end{array}
    \right)\; . 
  \end{equation}
\bigskip

The RRPA matrices A and B are composed of matrix elements of 
the residual interaction $V$, derived from an effective Lagrangian
density with medium-dependent meson-nucleon couplings 
\cite{NVFR.02}, as well as certain combinations of 
thermal occupation factors $n_k$. The diagonal matrix elements
contain differences of single particle energies $\epsilon_m-\epsilon_i$,
where $\epsilon_i < \epsilon_m$. Because of finite temperature, the 
configuration space includes particle-hole ($ph$), 
particle-particle ($pp$), and hole-hole ($hh$) pairs.
In addition to configurations of positive energy, the FTRRPA configuration
space must also contain pair-configurations formed from the fully
or partially occupied states of positive energy and the empty
negative-energy states from the Dirac sea. The inclusion of
configurations built from occupied positive-energy states and
empty negative-energy states is essential for current conservation,
decoupling of spurious states, and for a quantitative description of 
excitation energies of giant resonances~\cite{PRNV.03}.

The residual interaction is derived from the 
DD-ME2 effective Lagrangian \cite{LNVR.05},
and single-particle occupation factors at finite temperature are included
in a consistent way both in the FTRMF and FTRRPA. The same
interaction is used both in the FTRMF equations that determine the single-nucleon
basis, and in the matrix equations of the FTRRPA. The full set of
FTRRPA equations is solved by diagonalization. The result are excitation 
energies $E_{\nu}$, and the corresponding forward- and
backward-going amplitudes $X^{\nu}$ and $Y^{\nu}$,
respectively, that are used to evaluate the transition strength for 
a multipole operator $Q_J$:
\begin{equation}
   B(J,E_\nu) = \left | \sum_{mi} (X^{\nu,J}_{mi} + 
   (-1)^J Y^{\nu,J}_{mi} ) \langle m || Q_J || i\rangle (n_i
   -n_m) \right |^2.
\end{equation}
Discrete spectra are averaged with a Lorentzian distribution of arbitrary 
width ($\Gamma = 1$ MeV in the present calculation).

\begin{figure}
\centering
\includegraphics[scale=0.8]{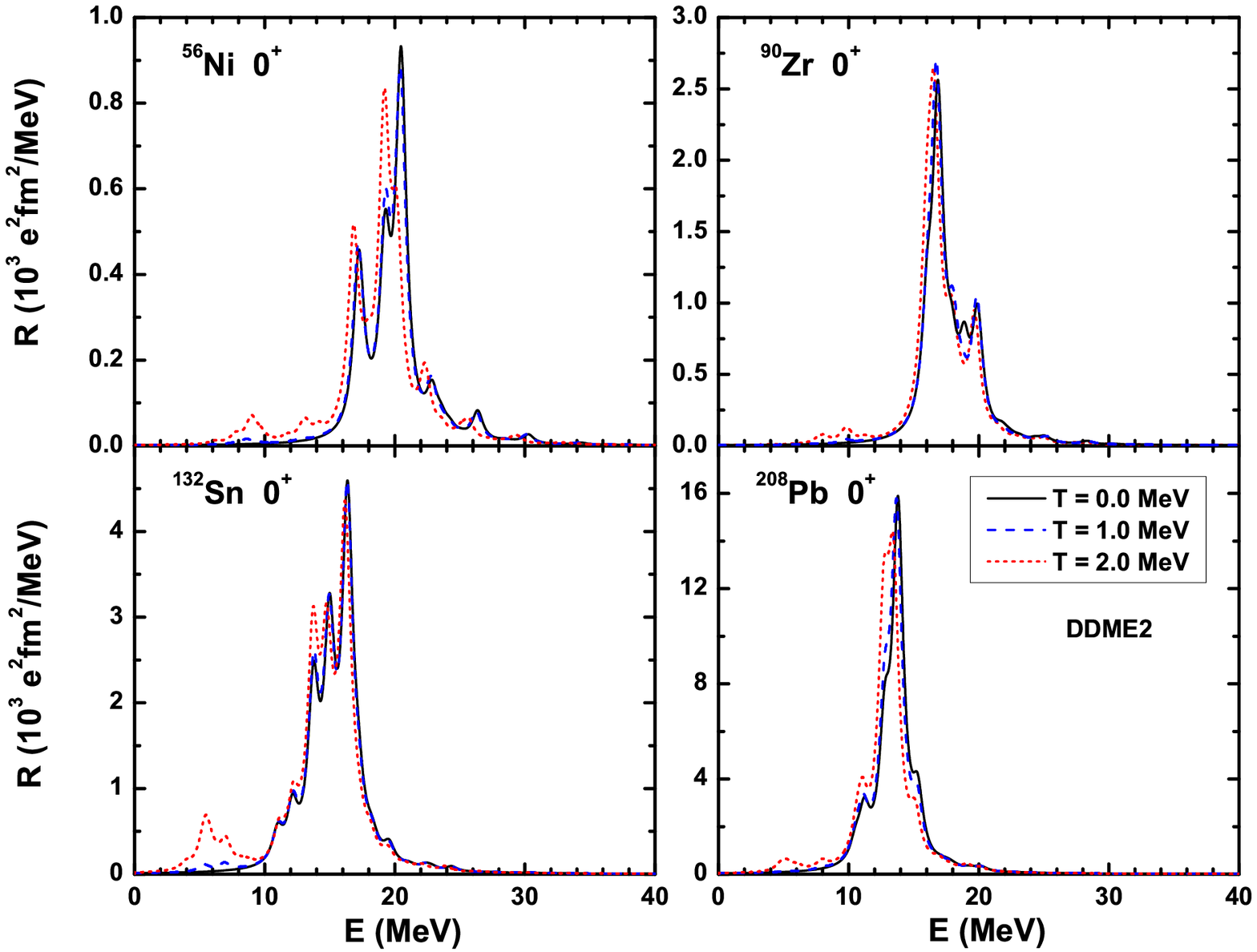}
\caption{Isoscalar monopole transition strength distributions
in $^{56}$Ni, $^{90}$Zr, $^{132}$Sn, and $^{208}$Pb, calculated
with the FTRRPA (DD-ME2) at temperatures $T=0, 1$, and 2 MeV.}
\label{monopole}
\end{figure}

In Fig.~\ref{monopole} we display the FTRRPA strength distributions in 
$^{56}$Ni, $^{90}$Zr, $^{132}$Sn, and $^{208}$Pb  for the isoscalar monopole 
operator, at temperatures $T=0, 1$, and 2 MeV. 
At zero temperature the transition strength distributions are clearly 
dominated by the pronounced collective isoscalar giant monopole
resonance (ISGMR). Even by increasing the temperature to 
$T=2$ MeV, the ISGMR gets only slightly modified in $^{90}$Zr, $^{132}$Sn, 
and $^{208}$Pb. In the lighter nucleus $^{56}$Ni, where
the ISGMR is less collective and therefore
more sensitive to modifications of the corresponding configuration 
space at finite temperature, at $T=2$ MeV the overall strength distribution 
is lowered by $\approx 1$ MeV. It is interesting to note that for all nuclei 
under consideration, in the temperature range $T=1-2$ MeV additional
transition strength appears at energies below 10 MeV. 
These low-energy transitions occur because of thermal 
unblocking of single-particle orbitals close to the Fermi level. The effect
of finite temperature is especially pronounced for
the neutron-rich nucleus $^{132}$Sn, due to transitions from
thermally populated weakly-bound neutron orbits.  
The low-energy strength is dominated by two rather
pronounced peaks. The state at 5.45 MeV corresponds to the 
single-neutron transition from thermally unblocked
orbitals:  3p$_{3/2}$  $\rightarrow$ 4p$_{3/2}$, whereas for the state at 
7.02 MeV the dominant transition is between the neutron orbitals
2f$_{7/2}$ and 3f$_{7/2}$.

\begin{figure}
\centering
\includegraphics[scale=0.8]{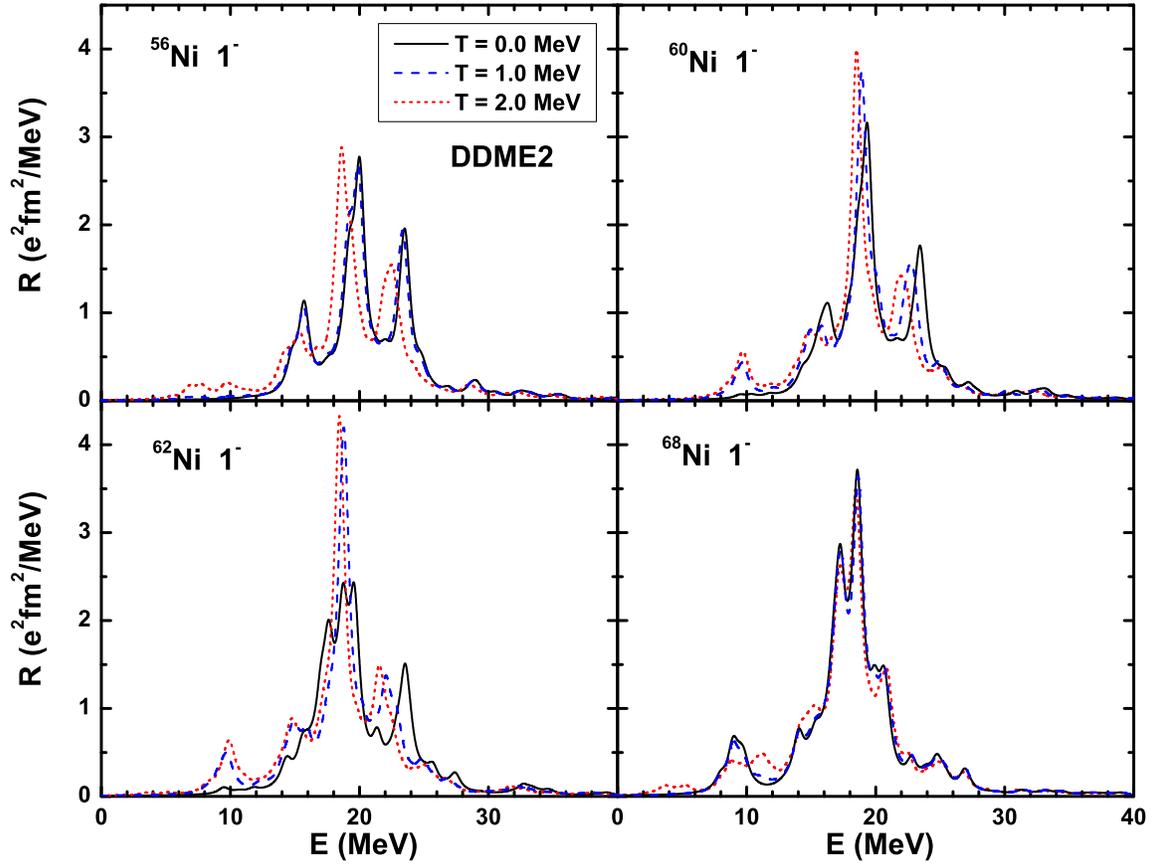}
\caption{Isovector dipole strength distributions in Ni isotopes 
at temperatures $T=0, 1$, and 2 MeV, calculated with the FTRRPA (DD-ME2).}
\label{nidipole}
\end{figure}

As already emphasized in the introduction, particularly relevant for
possible astrophysical applications is a microscopic description
of low-energy dipole response in hot nuclei. Fig.~\ref{nidipole} displays
the FTRRPA (DD-ME2) dipole strength distributions in Ni isotopes,
at temperatures $T=0, 1$, and 2 MeV. The transition
matrix elements are evaluated for the isovector dipole operator.
At zero temperature the main peak of the isovector giant dipole 
resonance (IVGDR) is located at $\approx$~18 MeV. 
For $^{68}$Ni, in particular, additional low-energy PDR structure
is predicted below 10 MeV even at zero temperature, and this result
was recently confirmed in the experimental study 
reported in Ref.~\cite{Wie.09}. The IVGDR is somewhat modified at 
finite temperature. At  $T=2$ MeV the main peak is lowered 
by $\approx 2$ MeV in $^{56}$Ni, whereas in $^{62}$Ni finite
temperature reduces the fragmentation of the IVGDR strength
in the interval $\approx$17-20 MeV, and enhances the collectivity 
of the main resonance peak. 

A very interesting finite temperature effect is predicted for 
the dipole strength distributions of $^{60,62}$Ni, 
where no low-energy excitations are present at zero temperature. 
Already at $T=1$ MeV novel transition strength develops 
below 10 MeV. For $^{60}$Ni, in particular, at $T=2$ MeV the main
low-energy peak is located at 9.71 MeV, it
exhausts 1.54\% of the Thomas Reiche Kuhn(TRK) sum rule, and 
is composed of 5 neutron and 4 proton single-particle
transitions:
$\nu 2p_{3/2} \rightarrow \nu 2d_{5/2}$ (45.87\%),
$\nu 1f_{5/2} \rightarrow \nu 2d_{3/2}$ (13.92\%),
$\nu 2p_{3/2} \rightarrow \nu 3s_{1/2}$ (8.43\%),
$\nu 1f_{7/2} \rightarrow \nu 1g_{9/2}$ (3.58\%),
$\nu 2p_{1/2} \rightarrow \nu 2d_{3/2}$ (1.30\%),
$\pi 2p_{3/2} \rightarrow \pi 2d_{3/2}$ (7.89\%),
$\pi 2p_{3/2} \rightarrow \pi 2d_{5/2}$ (6.32\%),
$\pi 1f_{5/2} \rightarrow \pi 2d_{3/2}$ (3.62\%), and
$\pi 1f_{7/2} \rightarrow \pi 1g_{9/2}$ (1.21\%).
The percentage in brackets corresponds to the contribution
of a particular transition to the total FTRRPA amplitude 
$\sum_{mi}(X_{mi}^2-Y_{mi}^2)(n_i-n_m)$ for the state at 9.71 MeV. 
This rather rich RPA structure shows that the new low-energy
state accumulates a small degree of collectivity due to thermal
population of both neutron and proton single-particle
levels around the Fermi surface. 

In the case of $^{62}$Ni, two pronounced low-energy peaks are calculated at  
temperature $T=2$ MeV: the states at 9.78 MeV and 10.03 MeV, 
exhausting 1.2\% and 1.5\% of the TRK sum rule, respectively. It is interesting to 
note that both states are dominated by proton transitions. For the state at 9.78 MeV: 
$\pi 2p_{3/2} \rightarrow \pi 2d_{5/2}$ (48.02\%),
$\pi 1f_{5/2} \rightarrow \pi 2d_{3/2}$ (20.99\%), 
$\pi 2p_{3/2} \rightarrow \pi 3s_{1/2}$ (3.12\%), 
$\nu 2p_{3/2} \rightarrow \nu 2d_{5/2}$ (17.06\%),  
$\nu 1f_{5/2} \rightarrow \nu 2d_{3/2}$ (4.36\%), and 
$\nu 1f_{7/2} \rightarrow \nu 1g_{9/2}$ (1.22\%).
The structure of the state at 10.03 MeV appears
to be less distributed, i.e. it is dominated by the transitions
$\nu 2p_{3/2} \rightarrow \nu 2d_{5/2}$ (26.55\%),
$\nu 2p_{3/2} \rightarrow \nu 3s_{1/2}$ (3.59\%),
$\pi 1f_{5/2} \rightarrow \pi 2d_{3/2}$ (63.60\%), and
$\pi 2p_{3/2} \rightarrow \pi 2d_{5/2}$ (2.55\%).

In $^{68}$Ni, where the PDR structure is present already at $T=0$,
the increase in temperature leads to fragmentation of the PDR 
and its spreading to even lower energy. These examples nicely illustrate the 
variety of effects one can expect for the dipole response of hot nuclei, e.g. the 
appearance of novel modes of low-energy excitations dominated by proton 
transitions, and the thermal spreading of PDR structures in neutron-rich nuclei.

\begin{figure}
\centering
\includegraphics[scale=0.8]{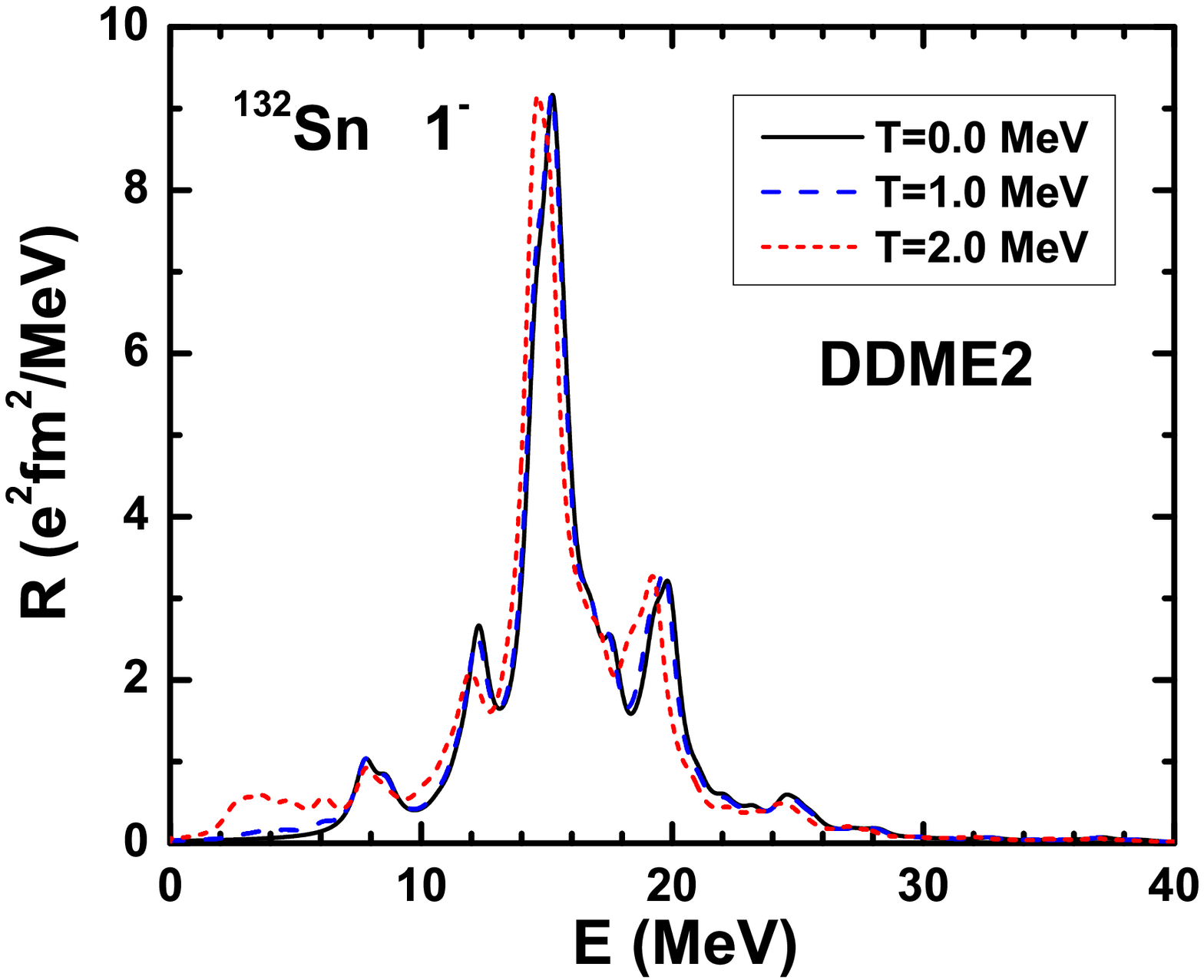}
\caption{Evolution of electric dipole transition strength with temperature in $^{132}$Sn, 
calculated with the FTRRPA (DD-ME2).}
\label{sndipole}
\end{figure}

Finally, Fig.~\ref{sndipole} shows the evolution of FTRRPA
dipole transition strength with temperature for $^{132}$Sn. 
This is an important example, because for this nucleus the 
occurrence of PDR was predicted in several theoretical
studies~\cite{PVKC.07}, and recently confirmed in Coulomb dissociation
experiments \cite{Adr.05}. With the increase of temperature up to 
$T \approx 1$ MeV the transition strength is virtually unchanged. 
Higher temperatures, however, have a pronounced effect
on low-energy dipole strength. At temperatures $\approx 2$ MeV, 
even though the main PDR peak at $E \approx 8$ MeV does not change 
much, additional low-energy single-particle transitions appear in 
the interval $E \approx 2-7$ MeV: 
E=3.75 MeV, $\nu 3p_{3/2} \rightarrow \nu 3d_{5/2}$ (99.65\%);
E=4.54 MeV, $\nu 2f_{7/2} \rightarrow \nu 3d_{5/2}$ (97.49\%);
E=4.92 MeV, $\nu 2f_{5/2} \rightarrow \nu 2g_{7/2}$ (99.58\%);
E=6.12 MeV, $\nu 2f_{7/2} \rightarrow \nu 2g_{9/2}$ (98.35\%).
At zero temperature the PDR state is calculated
at $E = 7.75$ MeV, whereas at $T=2$ MeV 
the PDR excitation energy is $E = 7.77$ MeV.
In Table~1 we compare the structure of FTRRPA states for the 
main PDR peaks at temperatures $T=0$ MeV and $T=2$ MeV. 
The amplitudes of transitions dominant at $T=0$ MeV get 
slightly reduced at higher temperature. Because of thermal unblocking of
the $\nu 1g_{9/2}$ orbit, at $T=2$ MeV an additional non-negligible 
contribution to the PDR peak corresponds to the transition 
$\nu$1g$_{9/2}$ $\rightarrow$ $\nu$1h$_{11/2}$.
\begin{table}
\begin{center}
\item[]\begin{tabular}{@{}*{3}{c}}
\hline\hline
  & $T=0$ MeV & $T=2$ MeV \cr
  & $E=7.75$ MeV & $E=7.77$ MeV \cr
\hline
$\nu 3s_{1/2} \rightarrow 3p_{3/2}$  &   51.85    &        50.37 \cr
$\nu 2d_{3/2} \rightarrow 3p_{3/2}$  &   19.15    &        17.22 \cr
$\nu 2d_{3/2} \rightarrow 3p_{1/2}$  &   11.56    &        10.43 \cr
$\nu 3s_{1/2} \rightarrow 3p_{1/2}$  &    6.64    &         5.48 \cr
$\nu 1h_{11/2} \rightarrow 1i_{13/2}$ &    4.99    &         4.44 \cr
$\pi 1g_{9/2} \rightarrow 1h_{11/2}$  &    2.17    &         2.43 \cr
$\nu 1g_{9/2} \rightarrow 1h_{11/2}$  &      -     &         1.26 \cr
\hline\hline
\end{tabular}
\caption{FTRRPA transition amplitudes for the main PDR peaks 
in $^{132}$Sn at
temperatures $T=0$ MeV and $T=2$ MeV. Included are 
the contributions (in \%) of dominant configuration to the total sum of FTRRPA
amplitudes: $\sum_{mi}(X_{mi}^2-Y_{mi}^2)(n_i-n_m)$.}
\end{center}
\label{table1}
\end{table}

The results for the evolution of dipole response with temperature (c.f. 
Figs.~\ref{nidipole} and \ref{sndipole}) show that the principal effect of 
finite temperature on low-energy strength that is already present at zero temperature, 
e.g. in $^{68}$Ni and $^{132}$Sn, is the spreading of this structure to even 
lower energy and the appearance of states that correspond to
thermally unblocked transitions.
This could be important for calculation of astrophysical neutron 
capture rates relevant for r-process nucleosynthesis, because these reactions
take place at finite temperature. As recently pointed out in Ref.~\cite{Lit.09},
neutron capture rates computed with the Hauser-Feshbach model 
are sensitive to the fine structure of low-energy dipole transitions, 
emphasizing the importance of reliable microscopic predictions for 
the evolution of PDR structures both with neutron number and nuclear 
temperature.

This work was supported by the Unity through 
Knowledge Fund (UKF Grant No. 17/08) and MZOS - project 1191005-1010. Y. F. Niu 
acknowledges support from the Croatian National Foundation for Science, Higher Education and
Technological Development. The work of J.M and D.V. was 
supported in part by the Chinese-Croatian project ``Nuclear structure far from stability". 

\newpage


\begin{thebibliography}{00}


\bibitem{PVKC.07} N. Paar, D. Vretenar, E. Khan, and G. Col\`o,
Rep. Prog. Phys. 70 (2007) 691
\bibitem{Sav.06} D. Savran, M. Babilon, A. M. van den Berg, M. N. Harakeh, J. Hasper,
A. Matic, H. J. W{\"o}rtche, and A. Zilges, Phys. Rev. Lett. 97 (2006) 172502
\bibitem{Sch.08} R. Schwengner, G. Rusev, N. Tsoneva, et al., Phys. Rev. C 78 (2008) 064314
\bibitem{Oze.08} B. {\"O}zel, J. Enders, H. Lenske, et al., arXiv:0901.2443v1
\bibitem{Adr.05} P. Adrich, A. Klimkiewicz, M. Fallot, et al., 
Phys. Rev. Lett. 95 (2005) 132501
\bibitem{Wie.09} O. Wieland, A. Bracco, F. Camera, et. al., 
Phys. Rev. Lett. 102 (2009) 092502
\bibitem{Pie.06} J. Piekarewicz, Phys. Rev. C 73 (2006) 044325
\bibitem{Kli.07} A. Klimkiewicz et al., Phys. Rev. C 76 (2007) 051603(R)
\bibitem{Gor.98} S. Goriely, Phys. Lett. B 436 (1998) 10
\bibitem{Lit.09} E. Litvinova et al., Nucl. Phys. A 823 (2009) 26
\bibitem{Gor.02} S. Goriely, E. Khan, Nucl. Phys. A 706 (2002) 217
\bibitem{Civ.84} O. Civitarese, R. A. Broglia, and C. H. Dasso, Ann. Phys. 156 (1984) 142
\bibitem{Vau.84} D. Vautherin and N. Vinh Mau, Nucl. Phys. A 422 (1984) 140
\bibitem{Bes.84} W. Besold, P.-G. Reinhard, and C. Toepffer, Nucl. Phys. A 431 (1984) 1
\bibitem{Fab.83} M. E. Faber, J. L. Egido, and P. Ring, Phys. Lett. B 127 (1983) 5
\bibitem{Rin.84} P. Ring, L. M. Robledo, J. L. Egido, and M. Faber,
Nucl. Phys. A 419 (1984) 261
\bibitem{Sag.84} H. Sagawa and G. F. Bertsch, Phys. Lett. B 146 (1984) 138
\bibitem{Lac.98} D. Lacroix, P. Chomaz, and S. Ayik, Phys. Rev. C 58 (1998) 2154
\bibitem{Gal.85} M. Gallardo, M. Diebel, T. Dossing, and R. A. Broglia, Nucl. Phys. A 443 (1985) 415
\bibitem{Bor.86} P. F. Bortignon, R. A. Broglia, G. F. Bertsch, and J. Pacheco, Nucl. Phys. A 460 (1986) 149
\bibitem{Dan.96} N. D. Dang, Phys. Rep. 264 (1996) 123
\bibitem{OBA.96} W. E. Ormand, P. F. Bortignon, and R. A. Broglia, Phys. Rev. Lett. 77 (1996) 607
\bibitem{Dan.97} Nguyen Dinh Dang and Fumihiko Sakata, Phys. Rev. C 55 (1997) 2872
\bibitem{Lar.99} A. B. Larionov, M. Cabibbo, V Baran, and M. Di Toro,
Nucl. Phys. A 648 (1999) 157
\bibitem{Sto.04} A. N. Storozhenko, A. I. Vdovin, A. Ventura, and A. I. Blokhin,
Phys. Rev. C 69 (2004) 064320
\bibitem{Mey.83} J. Meyer, P. Quentin, and M. Brack,
PHys. Lett. B 133 (1983) 279
\bibitem{Bar.85} M. Barranco, A. Polls, and J. Martorell, Nucl. Phys. A 444 (1985) 445
\bibitem{Goo.81} A. L. Goodman, Nucl. Phys. A 352 (1981) 30
\bibitem{Som.83} H. M. Sommermann, Ann. Phys. 151 (1983) 163
\bibitem{Kha.04} E. Khan, Nguyen Van Giai, and M. Grasso, Nucl. Phys. A 731 (2004) 311
\bibitem{Bra.89} A. Bracco et al., Phys. Rev. Lett. 62 (1989) 2080
\bibitem{Tve.96} T. S. Tveter et al., Phys. Rev. Lett. 76 (1996) 1035
\bibitem{Pie.96} D. Pierroutsakou et al., Nucl. Phys. A 600 (1996) 131
\bibitem{Bau.98} T. Baumann et al., Nucl. Phys. A 635 (1998) 428
\bibitem{Gam.00} Y. K. Gambhir, J. P. Maharana, G. A. Lalazissis, C. P. Panos,
and P. Ring, Phys. Rev. C 62 (2000) 054610
\bibitem{LKR.97}G.A. Lalazissis, J. K\"onig, and P. Ring,
     Phys. Rev. C 55, (1997) 540
\bibitem{NVFR.02} T. Nik\v si\' c, D. Vretenar, P. Finelli, P. Ring,
        Phys. Rev. C 66 (2002) 024306
\bibitem{Bon.84} P. Bonche, S. Levit, and D. Vautherin, Nucl. Phys. A 427 (1984) 278
\bibitem{PRNV.03} N. Paar, P. Ring, T. Nik\v{s}i\'{c}, and D. Vretenar,
    Phys. Rev. C 67 (2003) 034312
\bibitem{LNVR.05} G. A. Lalazissis, T. Nik\v si\' c, D. Vretenar, and P. Ring,
Phys. Rev. C 71 (2005) 024312


\end{thebibliography}
\end{document}